# 國立清華大學

## 碩士論文

# 一個基於可退之稅證明共識演算法的雙向鏈結區塊鏈架構

# A Double-Linked Blockchain Approach Based on Proof-of-Refundable-Tax Consensus Algorithm

**系所:電機工程學系碩士班**

**學號姓名: 104061591 江政勳** (Zheng-Xun Jiang)

**指導教授: 蔡仁松教授** (Prof. Ren-Song Tsay)

**中華民國一百零八年七月**

# A Double-Linked Blockchain Approach Based on Proof-of-Refundable-Tax Consensus Algorithm

*Student: Zheng-Xun Jiang*

*Advisor: Prof. Ren-Song Tsay*

***Department of Electrical Engineering***

***National Tsing Hua University***

***Hsinchu, Taiwan***

***R.O.C***

*July 2019*

# 摘要

在本篇論文中，我們提出一個雙向鏈結(double-linked)的區塊鏈架構，此架構能夠改善區塊鏈的效能以及確保區塊的不分叉、保持一致的鏈結。另外，在提出基於可退之稅證明(proof-of-refundable-tax)的共識演算法，使我們的研究可以建構出可靠度、效能、公平性及穩定性都很高的區塊鏈運行流程。可退之稅證明採用可驗證隨機函數(verifiable random function)去取代傳統挖礦方式，可驗證隨機函數的概率與個別參與者累計的可退之稅成正比關係，會影響到未來選上區塊鏈維護者的機率。每個人可退之稅的累計多寡可以作為個別參與者活躍的程度而被記錄，因此可退之稅可以有效防範人頭攻擊(Sybil attacks)的發生。此外，區塊鏈的完成獎勵會從每個維護者累計的可退之稅中扣除，這使我們的區塊鏈系統保持穩定的財富分配且避免"富者越富"的問題。我們已經對提出來的架構及共識演算法進行了測試，結果非常有希望實現。

# Abstract

In this paper we propose a double-linked blockchain data structure that greatly improves blockchain performance and guarantees single chain with no forks. Additionally, with the proposed proof-of-refundable-tax (PoRT) consensus algorithm, our approach can construct highly reliable, efficient, fair and stable blockchain operations. The PoRT algorithm adopts a verifiable random function instead of mining to select future block maintainers with the probability proportional to each participant's personal refundable tax. The individual refundable tax serves as an index of the activeness of participation and hence PoRT can effectively prevent Sybil attacks. Also, with the block-completion reward deducted from each maintainer's refundable tax, our blockchain system maintains a stable wealth distribution and avoids the "rich become richer" problem. We have implemented the approach and tested with very promising results.

# I. Introduction

Since 2009, the blockchain technology invented by Satoshi Nakamoto [1] has been successfully applied to decentralized secure cryptocurrency Bitcoin transactions [2]. With the success of Bitcoin, the world recognizes the potential impact of the blockchain technology and many works have then attempted to further improve the technology and extend the idea to other applications. On the other hand, blockchain also creates many issues that need to be solved, such as unreliable forking problem, inefficient processing time, unfair maintainers' verification process and unstable distribution of incentive wealth.

A blockchain is a growing list of blocks and each block contains a cryptographic hash of the previous block, a timestamp, and transaction records. By design, a blockchain is resistant to modification of the transaction records and is openly verifiable. Typically, a blockchain is managed by a peer-to-peer network which collectively adheres to a protocol for inter-node communication and new block validation [3].

In fact, the first cryptography-based payment system was proposed by David Chaum in 1982 [4]. The idea did not fly due to lacking of a publicly trustable framework. Then Haber and Sornetta in 1991 described a cryptographically secured chain-of-blocks approach which had a trusted third party to sign documents and timestamps to avoid tampering [5]. Nakamoto [1] then extended the Hashcash proof-of-work method [6] to add blocks to the chain in a critical way that does not require the blocks to be signed by a trusted party. The Nakamoto proof-of-work (PoW) consensus scheme has served as a key component of the decentralized bitcoin blockchain.

The Nakamoto PoW scheme allows the majority of network nodes to reach an agreement on the public distributed ledger (or block) under Byzantine Fault Tolerance condition [7]. Specifically, for the Nakamoto PoW approach, any node can propose a new block to the blockchain by generating a valid random number (called nonce) which makes the block hash value smaller than a target value. This random number guessing work is often referred to as puzzle-solving. The probability of success to propose the first legal block is depending on how much work the node's computing power can crank out and that is why the puzzle-solving work is often named proof-of-work mining. Since every node can freely compete in the block mining, a critical issue of the PoW approach is that this scheme can waste huge amount of collective energy. Another issue is the uncertain finality of the PoW proposed block. Since a new block can link to any existing block, one can never be certain if a proposed block will be always extended by other blocks. In reality, multiple forks may occur.

Essentially, the success rate of the Nakamoto PoW mining task is proportional to the computing power a miner can spare. Therefore, some even invest hugely on the development of Application-Specific-Integrated-Circuits (ASICs) for mining tasks in order to win the race. Some collaborate and form mining pools to increase the winning probability and then share the reward. Nevertheless, the pooling or dominating computing force essentially goes against the decentralization principle of blockchain. Furthermore, if the network mining power of any party exceeds 51%, then the blockchain can even be controlled or privatized [8].

Therefore, some have proposed the proof-of-stake (PoS) algorithms to improve the issues of PoW by choosing maintainer to create the next block via random selection or wealth amount or age or combinations of a few different parameters [9]. Specifically, the Casper PoS is based on Dagger-Hashimoto [10] [11] ASIC-resistant algorithm for mining. Casper sets the mining puzzle's difficulty inversely proportional to the miner's stake in the network. A stake is essentially a locked account with a certain balance representing the miner's commitment to keep the network healthy. With higher stake, one can create a valid block using less computing power. Therefore, the PoS approach consumes less collective network computing energy and is more efficient.

However, the PoS scheme may result in centralization, as rich members have more advantage to commit new blocks, because they can put in higher stake. The centralization phenomena would cause an unreliable, unfair selection result and the rich will become richer.

In contrast to the randomness of mining winners in PoW and PoS, the delegated scheme (e.g. delegated proof-of-stake, DPoS) lets each shareholder delegate someone into the management board and the board members take turn in a round robin order to create blocks. The DPoS uses the reputation algorithms and real-time voting to select fixed delegates, typically around 100 selected delegates [12]. Practically, the DPoS miners, or known as delegates, can collaborate to deliver blocks and use PBFT-like verification to agree on a unique block. Because the number of the DPoS miners is smaller, in multiple order of magnitude, than that of the PoW or PoS consensus algorithms, therefore the DPoS exhibits higher verification efficiency.

However, the DPoS is more oriented for centralization than the PoW and the PoS because the number of delegates is limited. If more malicious stakeholders can accumulate reputation quickly, they may easily fill in the delegate quota. Then, the verification process will not be fair for loyal users. Obviously, the incentive reward will be unstably distributed when malicious stakeholders are easily selected in the long-term.

To have a fair selection of delegates, the Algorand [13] applies verifiable random functions (VRFs) [14] for participants to self-verify, broadcast and claim to be blockchain maintainers randomly in a private and non-interactive way. Algorand divides the maintainers into two different roles, creators and voters. By separating block creators from voters, the verification uses voters to vote for the largest claimed creators' hash value of block rather than the whole of block contents. Hence, the blockchain verification process can be performed more efficiently. However, due to the self-claim scheme, there are uncertain number of successors, so they need to be confirmed after rounds of PBFT-like voting and the process is relatively inefficient. Additionally, since the qualification probability is proportional to the participant's account balance, this approach can also result in rich become richer issue.

To alleviate the above issues, we propose a Proof-of-Refundable-Tax (PoRT) blockchain consensus scheme which separately selects block creators and voters using a new verifiable random selection scheme. The proposed scheme virtually forms a double linked chain and is resistant to monopolism. Every network participant can be selected as a block creator or voter with a probability proportional to the individual refundable tax, which is the collected fixed-rate transaction tax from the participant's past transactions. Once the selected maintainer successful completes the block creation or validation task, a certain amount of tax is refunded as a reward and deducted from the individual's total refundable tax. The PoRT scheme basically avoids the "the rich become richer" phenomenon by constantly resetting the selection base, i.e. the refundable tax, and hence is fair to all participants. Additionally, since the tax level reflects the activeness of a participant, active users have more incentive to ensure the working state of the network.

Most importantly, the proposed PoRT scheme forms a unique double-linked blockchain data structure which guarantees a single chain with no side branches. Essentially, the inclusion of the last block's hash value forms a backward link while the PoRT specific verifiable selected list of future block maintainers forms a forward link. The combination of backward and forward links establish a secure double linked blockchain structure.

Our proposed PoRT consensus scheme greatly improves the efficiency, reliability, stability and fairness over existing blockchain consensus algorithms. Our approach is very efficient as it requires no costly mining computations. The PoRT approach is fair as it utilizes a verifiable random hash value deterministically computed from the current block data and the current block's unique address of each maintainer. The randomized hash numbers corresponding to current block maintainers are used to select the successors impartially and verifiably with a probability proportional to individual's refundable tax. Our scheme is stable as the block creators and voters are refunded with their own paid taxes. Once tax is refunded, the refundable tax is reduced. Therefore, one has less probability to be repeatedly selected. This approach is also highly reliable as it guarantees a single chain with no side branch due to the double-linked structure. Therefore, double-spending issue simply cannot occur and PoRT blockchain finality is almost instant.

The PoRT blockchain consensus algorithm and double-linked data structure can effectively solve the fairness, stability, reliability and efficiency issues existing in current approaches. We will elaborate details of our proposed approach next in the paper.

The remainder of this paper is organized as follows. Section 2 reviews related work. Section 3 presents details of the proposed PoRT consensus algorithm. Section 4 discusses the algorithm design. Section 5 propose some future work with need to be concerned and finally we briefly conclude the paper in Section 6.

# II. Related work

In the following, we review some representative blockchain consensus algorithms in more details.

**Proof-of-work (PoW).** Bitcoin is the first publicly accepted cryptocurrency based on the proof-of-work (PoW), or also called the "Nakamoto consensus" algorithm [13]. The PoW algorithm requires solving a cryptographic puzzle for which one needs to compute a nonce such that the cryptographic hash value is less than a pre-set target value. The first who finds a nonce solving the puzzle will be the leader to propose a block. Since everyone can involve in solving puzzles but only one wins, the Bitcoin PoW process is inefficient and may waste computing, energy and equipment resources.

Another issue of PoW is that everyone can compete to build his own chain, therefore multiple side branches can exist in parallel and hence theoretically no finality can be reached. Practically, Bitcoin forces confirmation of the proposed block after six continuous blocks. The existence of side branches or forks results in unreliable blockchain records, as hackers can deliberately create double spending transactions on different branches.

Additionally, one may equip with high-power ASIC mining machine or form a collaborative mining pool to gain unfair advantage of winning reward through puzzle-solving or block creation. Therefore, the wealth distribution eventually will be dominated by a few high-power participants and is hence unstable.

**Proof-of-Stake.** Ethereum proposed Ethash to avoid "ASIC equipment race" through Dagger-Hashimoto method which requires memory-access I/O to restrict ASIC performance [15]. Furthermore, Ethereum proposed Casper proof-of-stake algorithm to replace the proof-of-work algorithm. Participants in Casper deposit their stakes (i.e. monetary value or amount or age etc.) to reduce their computing burden of finding-puzzles in the network. The stake is essentially a locked account with certain balance representing the miner's commitment to keep the network healthy [16]. Hence, the more stake a participant deposits, the puzzle difficulty is lower. However, a malicious block creator may purposely create multiple forks with no additional stake but higher probability to succeed. This phenomenon is called "nothing-at-stake," which results in unreliable risks for new blocks. Ethereum prevents "nothing-at-stake" [17] through a "detect and punish" mechanism which catches malicious creators and deprives their stakes. But, in general, the depositing stake and the burden of punishment are not small overhead for everyone. The richer participants in system would slowly become the only ones who can support and maintain the proof-of-stake system. The system's rewards can easily go to the richer ones and result in "the rich become richer" and centralization phenomenon. Centralization generally is more efficient but loses fairness for others to have equal opportunity to participate in the system. Obviously, the biased reward distribution in Ethereum would impact the stability of the account balance distribution of the whole network.

**Delegated proof-of-stake (DPoS).** The delegated proof-of-stake algorithm lets each shareholder delegate someone into the management board and the board members take turn in a round robin order to create blocks. Examples include EOS [18], BitShares [12] and Lisk [9]. The process is first developed by Daniel Larimer [12] as an alternative to energy-inefficient proof-of-work and proof-of-stake algorithms. An active user holding tokens may vote for delegates. The top ranked token collectors become delegates to build the management board. The management board use the round-robin scheme to assign one to create block and others to validate block.

The DPoS management board can be counted as a centralized scheme, since the number of delegates usually is fixed at 21 to 101. Obviously, the limited board scale is more energy efficient for execution as compared with the whole-network scale of proof-of-work and proof-of-stake mining. However, under the current delegate voting scheme, active users tend to repeatedly vote for the same delegates who simply do the job. Hence, a small group of delegates may dominate the management board. Then the management would be unreliable as the small group of constant delegates may intentionally isolate certain transactions, reject blocks not of self-interests. Worse yet, malicious attackers may work to take over most delegate seats and dominate the block creation business and cause unfair blockchain participation. Consequently, the rewards may concentrate on a small group of delegates and cause unstable wealth distribution.

**Verifiable random functions (VRFs).** Instead, Dfinity [19] and Algorand [13] select at each round a random set of nodes to propose blocks. In Algorand, a block creator first generates a random seed by encrypting on the previous block's random seed using the creator's private key. The new random seed is used to compute a threshold probability value for each block maintaining role in a verifiable way. Then any network participant may self-qualify to be the new maintainer based on a probability proportional to current personal balance. The whole process is verifiable once the new block random seed is announced [13]. Therefore, an issue may occur if the block creator reveals the random seed before completing block creation, then adversaries may do targeted attacks on those who have high probability to be maintainers. Another issue is that, since Algorand also prefers candidates with higher account balances, like PoS, there are concerns of reliability and fairness issues as rich can become richer. However, the blockchain has less probability of forking.

One advantage of the self-verify scheme is that no one knows who are the maintainers and hence the approach reduces the possibility of targeted attacks or Sybil attacks. Additionally, Algorand can scale relatively easily to a large number of users using the VRF self-qualification scheme. Additionally, since the random seed is generated based on the previous block random seed and the creator's private key, hence the random seed is not a predictable, but anyone can validate using the creator's public key. Similarly, the lottery threshold values can also be validated but cannot be predicted. The self-qualified maintainers can immediately participate after they match the value and send a check-message.

The Algorand generates a dynamic set of creators and voters to maintain the blockchain while avoiding the technical puzzle-solving, so the efficiency is higher than the PoW and PoS approaches. With strong synchrony, that requires that most (more than 95%) honest users can send messages and be received by most others, the Algorand's Byzantine agreement will always pick only one block and create no fork. Therefore, the Algorand blockchain in strong synchrony is reliable with no double-spending problems.

Although everyone can fairly participate in the VRF-based self-qualification scheme, the approach prefers those with higher account balances, hence the associated rewards may gradually go to the rich users and cause unbalanced wealth distribution.

To resolve the above-mentioned issues, we modify the VRF approach and propose a double-linked blockchain data structure based on proof-of-refundable-tax consensus algorithm. Our approach has two key contributions. First, the design of double-linked data structure secures fork-less single chain. The traditional blockchain has only a backward link to previous block's hash value. The backward link structure does ensure immutability of the blockchain records. However, existence of forks could not guarantee finality. Whereas, our approach explicitly selects future maintainers and generates an implicit forward link and hence a unique double-linked blockchain structure which generates only fork-less single chain.

Secondly, our method is better in fairness and stability since block creation rewards are deducted from the refundable taxes to reset the probability of becoming maintainers. Those who have been maintainers will have less probability to be selected again. Therefore, the tax refund serves as a system damper and give everyone fair opportunity to win rewards and hence the distribution of the account balances will be stable. Essentially, users who do more transactions will pay more tax but also have more opportunity to get refund back. Therefore, in general the individual account should maintain a stable balance.

In summary, our proposed PoRT maintainer selection approach greatly enhances the efficiency and reliability and the double-linked data structure greatly enhances the fairness and stability. Next, we shall elaborate our proposed approach.

# III. The PoRT Consensus Algorithm and The Double-Linked Data Structure

Our proposed PoRT consensus algorithm greatly improves the efficiency, fairness, stability and reliability of blockchain operations. We also devise a unique double-linked blockchain data structure specifically to enhance the reliability and efficiency. In the following, we elaborate the key ideas.

## A. *Proof-of-refundable-tax Scheme*

The proposed PoRT scheme is a PBFT-like approach which selects block maintainers from candidates to create and validate blocks with consensus. Furthermore, PoRT divides the maintainers into two groups, the creators' and the voters' groups each with a specific duty. The creators are responsible for generating new blocks and the voters are responsible for validating the creators-generated blocks. Practically, a few creators and a few hundred voters are sufficient to guarantee *reliability*, therefore the proposed PoRT consensus scheme is very *efficient* as compared with other consensus algorithms that require every node in the network involves in block creation and validation.

As for the selection of maintainers, at the end of each consensus verification process, the PoRT algorithm conducts a verifiable selection process and picks the next group of maintainers. The selection process is a deterministic verifiable random mechanism based on each candidate's total refundable transaction tax and on-block generated public data, such as the block hash, maintainer's address and transaction records.

In order to avoid the unfair *rich become richer* issue occurred in a few other algorithms, after the completion of each block creation process, PoRT rewards each maintainer some work pay, which is then treated as a tax refund and is deducted from the corresponding individual refundable tax. Therefore, in the next selection processes these have-been-selected maintainers will be less likely to be selected again. The amount of work pay is dynamically adjusted to promote reasonable participation rate to ensure a secure consensus network.

The PoRT scheme is *reliable* from most network attacks. For instance, the refundable tax approach provides similar effectiveness as the Proof-of-Stake approach and can effectually prevent the Sybil attacks. Mainly this is due to the fact that PoRT basically uses the tax as an index of activeness of network participation for maintainer selection and sybils will not be qualified for selection. Additionally, multiple maintainers provide sufficient redundancy and effectively defend DDoS attacks.

To give an idea of the effectiveness of our approach, we assume each time we select 300 maintainers. For PBFT to work, there should be less than 1/3 faulty nodes. Therefore, we may assume the worst-case probability of a maintainer being faulty is $p = 1/3$. Then, for the 300 maintainers to fail Byzantine condition the worst-case probability is $\sum_{i=100}^{n=300} p^i(1-p)^{n-i} \approx 5.2 \cdot 10^{-53}$, a negligible number, when more than 100 maintainers are faulty.

Next, we shall discuss our proposed unique double-linked data structure that can provide instant finality and a fork-less blockchain.

## *B. Double-Linked Data Structure*

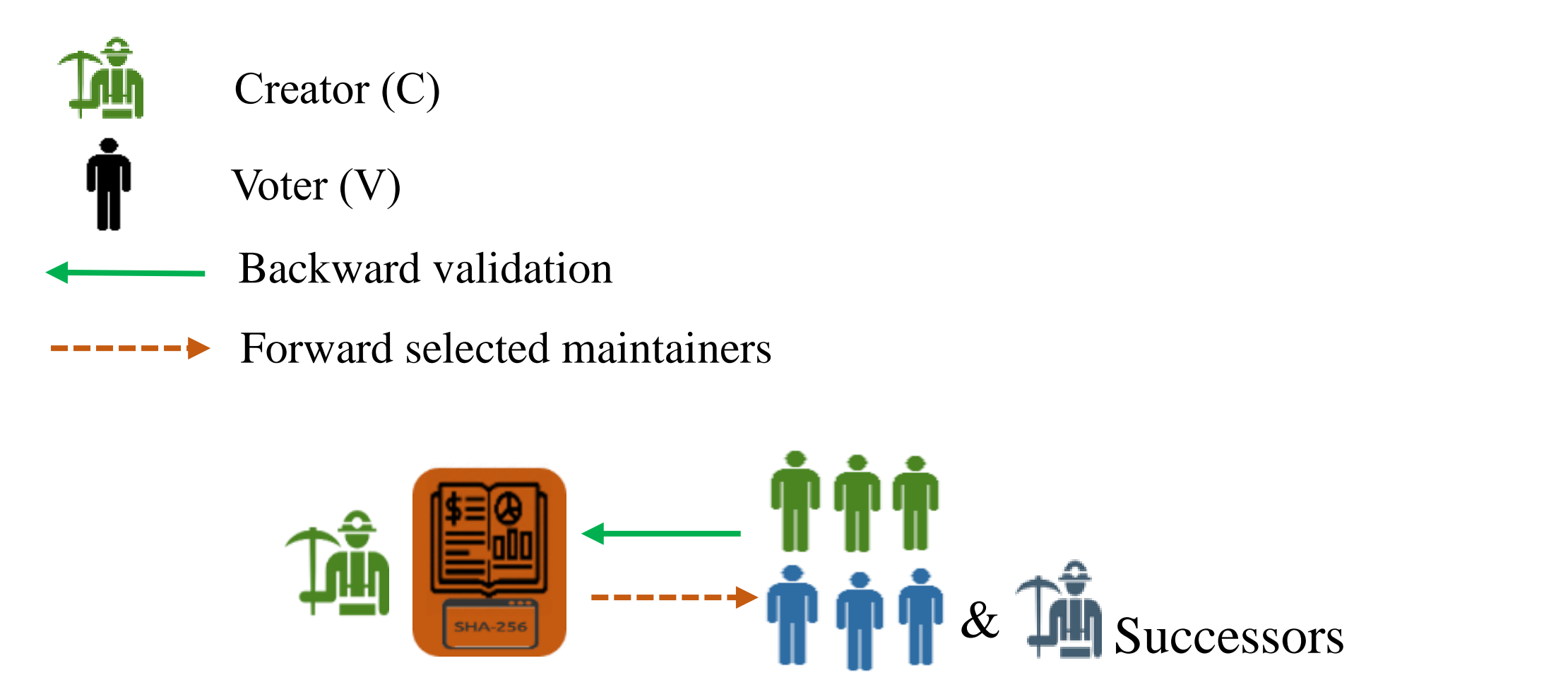

Figure 1: The PoRT approach establishes the forward link by selecting the inheriting maintainers for future blocks and the backward link by the hash pointer to the previous block. The forward and backward links form the unique PoRT double-linked structure.

Every blockchain design has a specific block data structure, which usually includes a back-linked hash pointer, a Merkle tree of transaction-related records.

PoRT creates a chain with no side branches or forks. In the PoRT scheme, the creator first collects a number of qualified unprocessed transactions and gathers sufficient validation votes to the last block and then generates a backward-linked hash pointer to the last block. Then with the verifiable current block hash value and the account address of each maintainer of the current block, the creator selects the inheriting maintainers of future blocks according to the refundable tax amount of each candidate. Since only the selected maintainers can create and vote in the future block processing, in this way PoRT establishes an implicit forward link to the future blocks.

Therefore, PoRT constructs a unique double-linked blockchain data structure.

As shown in Fig. 1, every creator-constructed block needs to be validated by voters assigned to the next block. Essentially, the voters assigned to the current block is to validate whether in the last block, the selection results, the voting results and the block construction are all valid and issue a signed approval vote if so. This check-balance act essentially prevents faulty creator. The key difference of our approach is that the votes of the voters is to commit the *last* block while the consensus standard follows the PBFT condition of 2/3 voters. With this high fault-tolerant standard, our double-linked blockchain structure guarantees the *reliability* of verification and selection results. Hence, the whole process is verifiable. Anyone can *efficiently* retrieve and verify any part of the blockchain.

We also devise an anti-collusion scheme in the PoRT approach. For a faulty block to be validated, a creator needs to collude with the voters. To break the collusion possibility, the key is to avoid having the creator and the voters generated from the same source. Next, we discuss our proposed anti-collusion approach.

## *C. The Jump-step-validation Anti-collusion Approach*

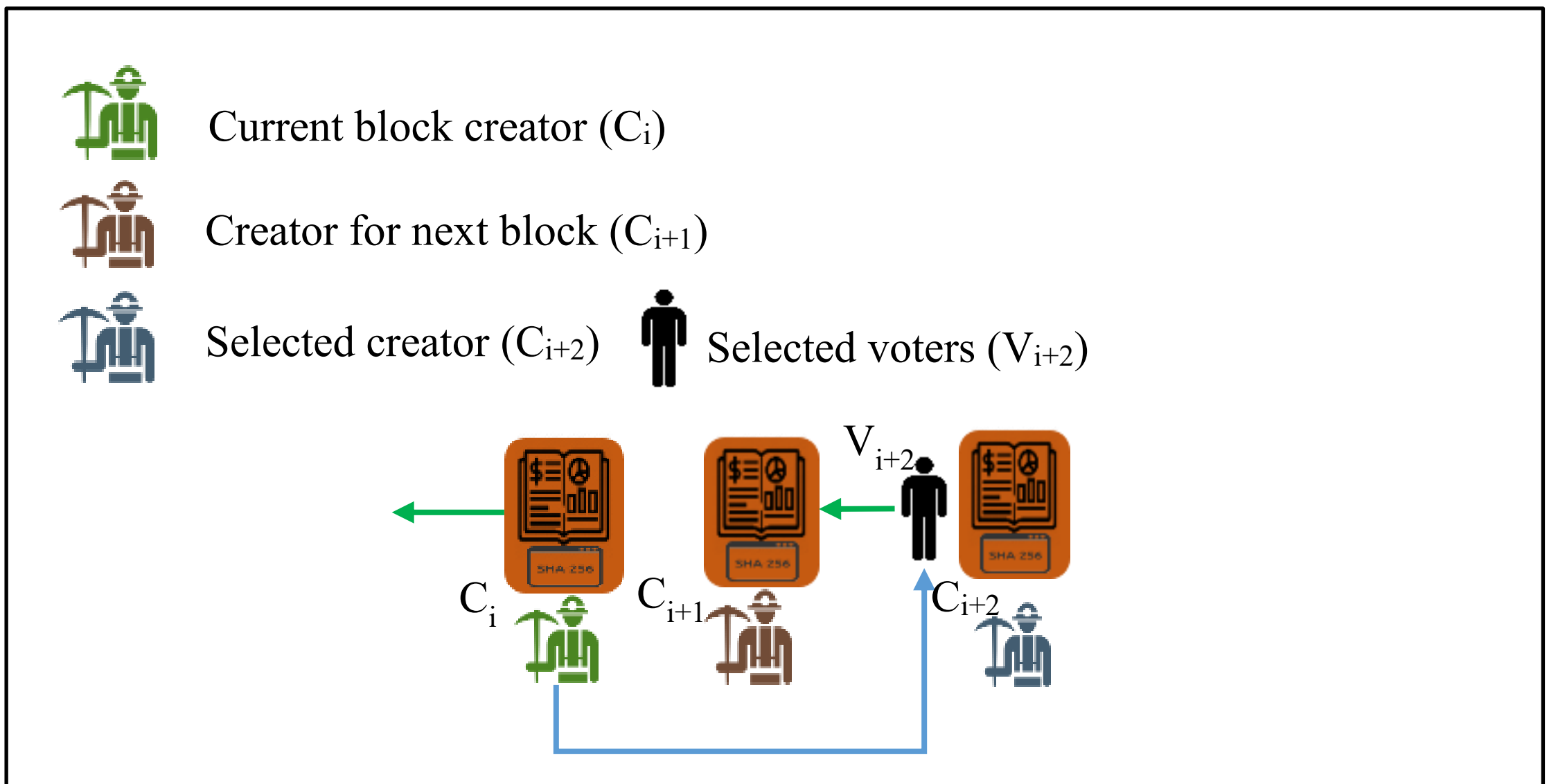

Figure 2: The proposed PoRT anti-collusion scheme has the selected creator to construct next-next block but the selected voters to validate the next block created.

The general principle of anti-collusion is to separate the block creator, the selected creators and the selected voters to avoid conflict of interest. PoRT applies the verifiable random selection method to select creators and voters independently and fairly and hence neutralizes the selection procedure. We then let the selected group of creators and the group of voters work on different blocks to avoid collusion. Essentially, the selected voters should not validate the current block, since they are selected by the current block creator. Similarly, the selected voters should not be used to validate the blocks created by the selected creators since they are from the same source.

Note that for the process in Fig. 1, a faulty block creator may intentionally assign a group of colluded inheritors. Since the selected creator and voters are colluded, the validity of the next creator created block cannot be guaranteed. In contrast, for the improved scheme shown in Fig. 2, we have the selected voters validate the next block while the selected creator is to create the next-next block. Since the selected maintainers are placed on the block after next block, or jump over next block, we name this procedure the jump-step validation.

For algorithm design, we modify the Merkle state tree to record the refundable tax amount and selection results. With the recorded results, one may easily verify the responsible creators and voters at each step. This jump-step validation method effectively avoids the possibility of creator-voters collusion. However, at each step if there is only one single creator, the approach may subject to crash failure.

## *D. Redundant Creators*

The PoRT consensus approach extends the possibility of Byzantine failure and crashing failure. To avoid failures, we simply have redundant blocks. As shown in Fig. 3, we have two creators to construct two redundant blocks $B_{i+1,1}$ and $B_{i+1,2}$ to reduce the block commit error. For example, if the creator of the block $B_{i+1,2}$ crashes, there are still one redundant block to be considered. In fact, we allow asynchronous synchronization and hence even if all crash, the system simply wait until at least one recovers to continue the process.

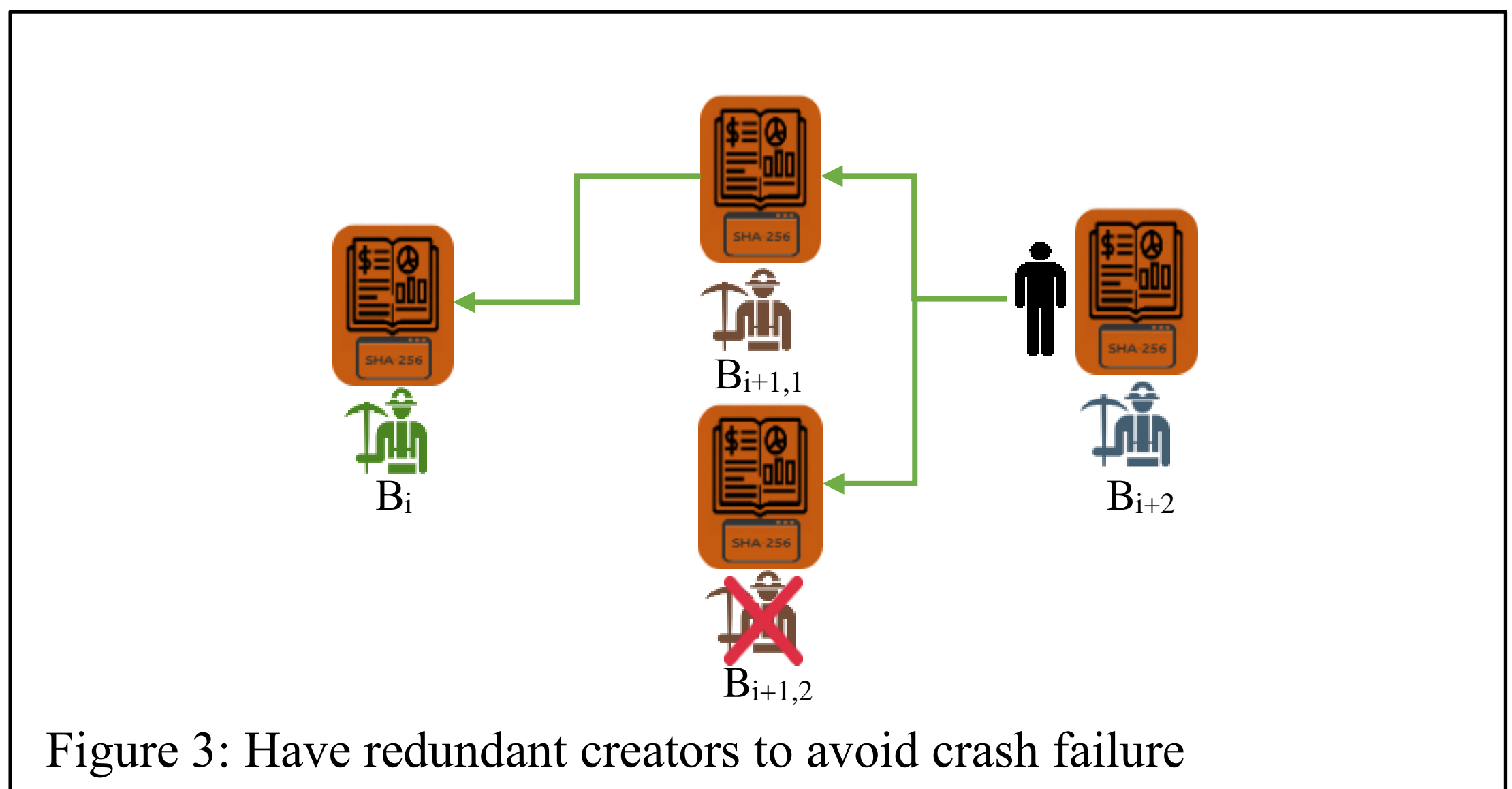

Figure 3: Have redundant creators to avoid crash failure

## *E. Penalty of Frauds*

Although the PoRT approach is equipped with many security mechanisms, just like any public blockchain it is subjected to potential attacks. We further devise a penalty scheme as an after-fact remedy in case any fraud is identified. Since any transaction has to be recorded on the blockchain to be effective, the validity can be thoroughly verified. The PoRT system allows anyone to report frauds to the block creator and if the fraud report is accepted and validated by the voters, then a reward is granted to the fraud reporter. At the same time, a penalty is applied to the convicted accounts. One way is to tag blacklist to the convicted node address in the Merkle state tree. The system then deprives the privilege of being the maintainer candidates from the blacklisted nodes for a certain period of time. If necessary, the system may confiscate the account balance of the convicted node.

# IV. Double-Linked and PoRT Consensus Algorithm Design

In this section, we elaborate double-linked and algorithm design details, discuss the data structure and consensus algorithm and how they achieve high reliability, efficiency, fairness and stability.

## *A.* Data Structure

### *1)* Merkle Patricia Tries

We adopt the Merkle Patricia tries used in Ethereum [15] which basically is a key-value map. The keys are addresses of users or smart contracts and the values are account states, which in Ethereum include the account "balance", "nonce", "code" and "storage" information. The accounts are fully deterministic, i.e. the same key is guaranteed to bind to the same value.

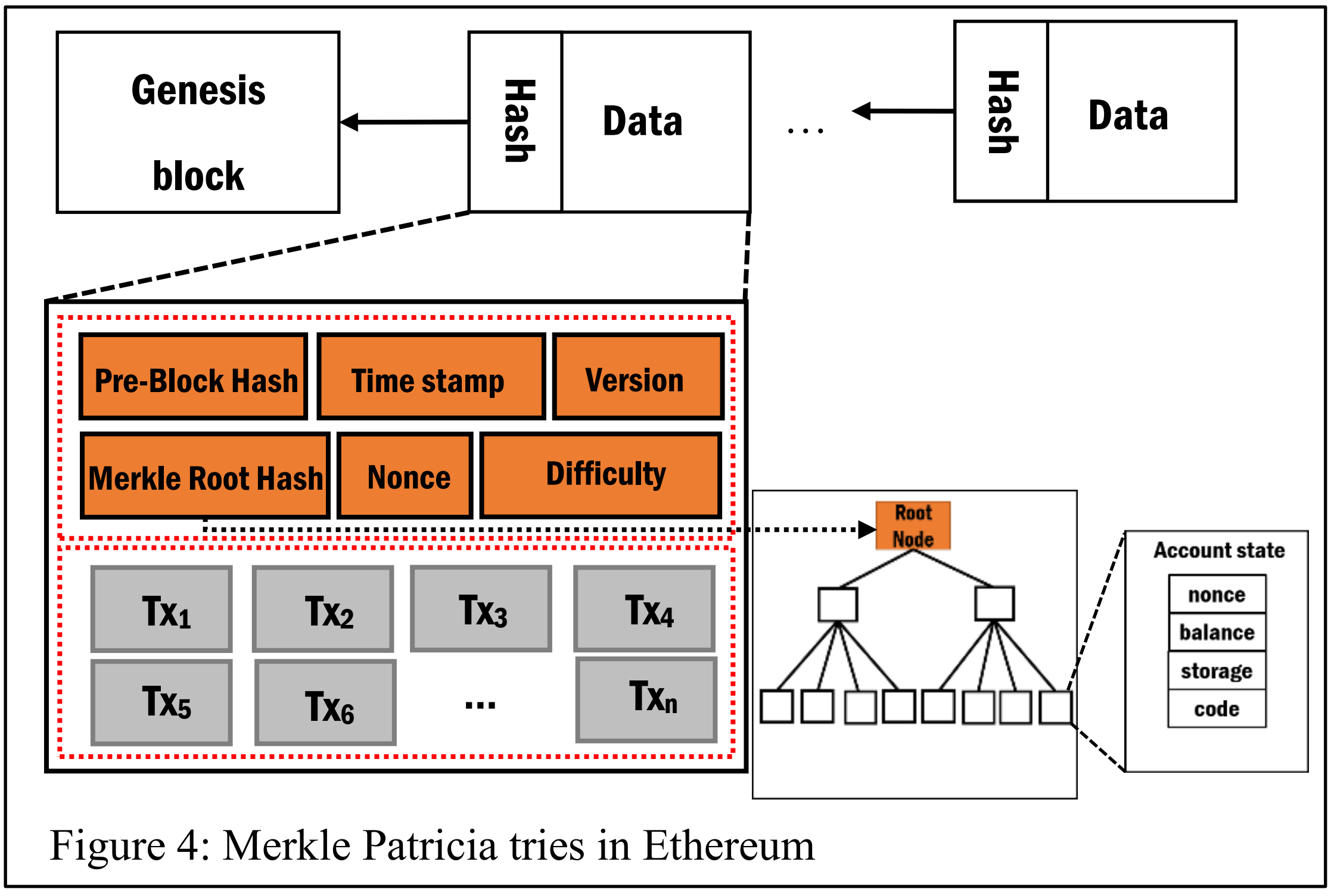

Figure 4: Merkle Patricia tries in Ethereum

### 2) *Modified Merkle Patricia Tries*

To record tax, selection result and penalty status, we add a few additional account state records, such as “tax”, “maintainer” and “blacklist” in the Merkle Patricia trie to facilitate PoRT implementation.

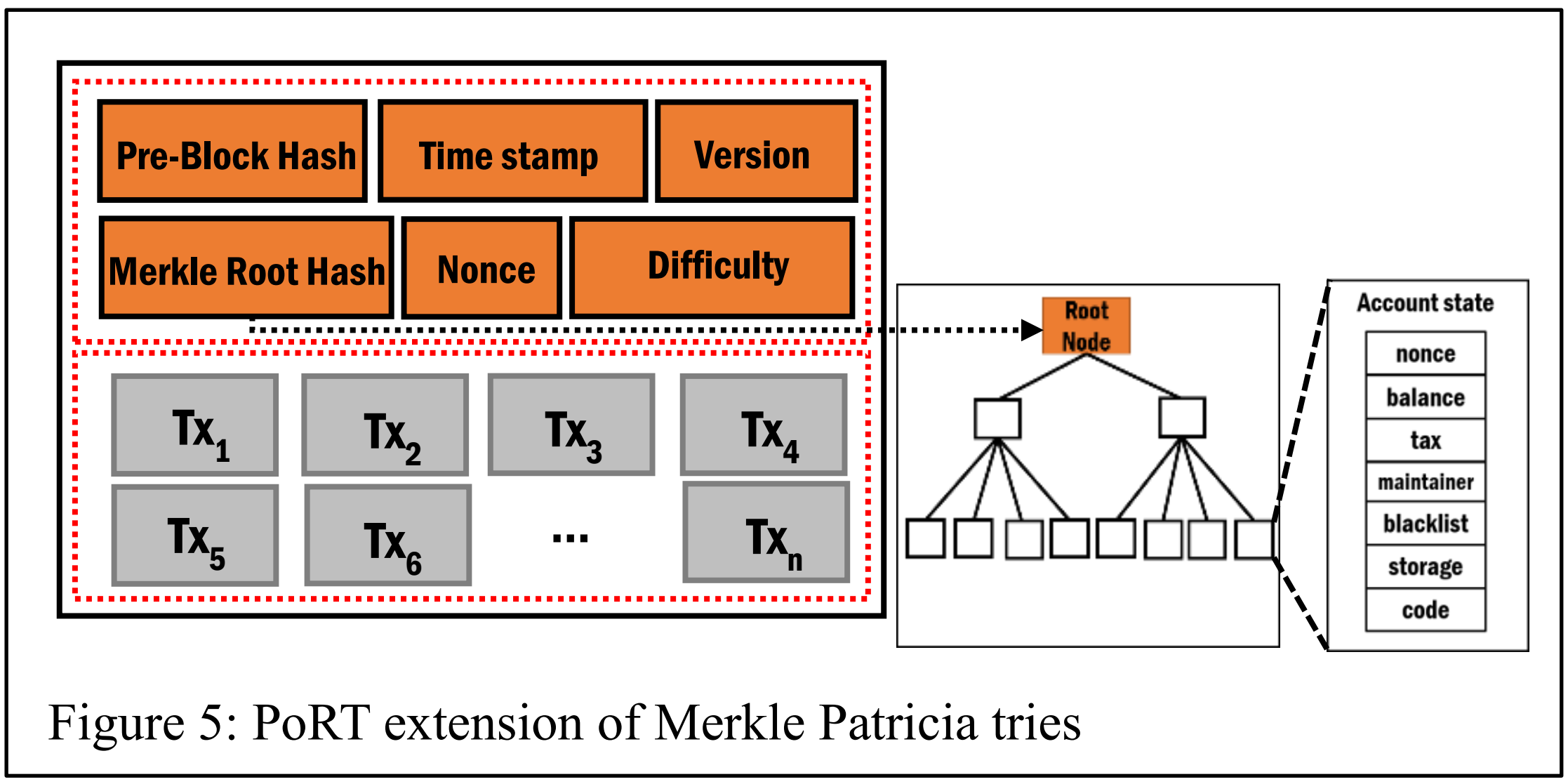

Figure 5: PoRT extension of Merkle Patricia tries

### 3) *Double-Linked Chain Structure*

The most unique design in the PoRT approach is that each block establishes both a backward link and a forward link. The unique double-linked design creates a secure single chain with no forks and hence can achieve instant finality. Each block records the previous block’s hash value which is a hash pointer to link to the previous block and serve as a backward link. The forward link is implied by the selected future block maintainers. Since only the previously elected maintainers can create and validate current block, a forward link is established and stored in the “maintainer” list in the modified Merkle Patricia tries. With no possibility of forking or double spending, the PoRT constructed chain is highly reliable.

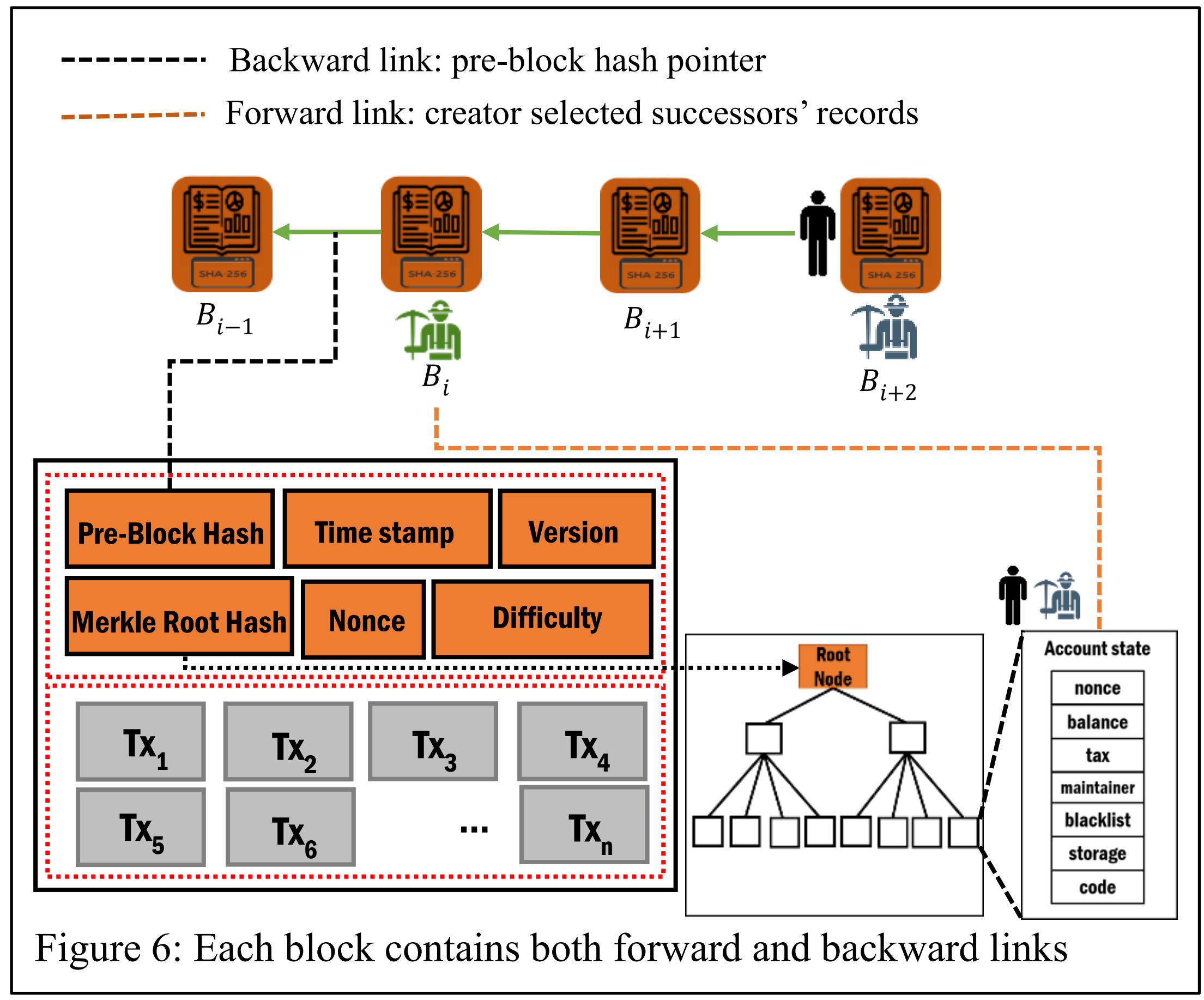

Figure 6: Each block contains both forward and backward links

In the next paragraph, we shall explain our PoRT consensus algorithm, which selects maintainers *fairly* and avoids the *rich become richer* issue. The algorithm is very *efficient* as only the selected maintainers are involved in creating and validating the blocks and no mining is required.

## *B.* Consensus Algorithm

Our proposed proof-of-refundable-tax (PoRT) algorithm uses "tax" as an index of user activeness. In the PoRT blockchain, both the senders and the receivers of a transaction need to pay a certain percentage of the transaction value as the transaction tax. Therefore, the more tax a user paid, the more active a user is. Intuitively, these active users have more incentive to maintain a working system. Also, the tax-based selection process can defend Sybil attacks since artificial accounts cannot accumulate enough tax to participate in the selection process.

The PoRT process is further divided into selection and verification steps, both require the collaboration of creators and voters to achieve consensus. After collected transactions for block creation are verified, the creator uses the block hash, each current maintainer' address and order to generate a hash value which serves as a random number to select the inheritor of the maintainer according to the ratio of the user's tax. This selection process can be calculated and verified by anyone.

Note that in our approach we have redundant creators to enhance reliability. Therefore, at the same time voters needed to verify previously created redundant blocks. The current block creator then collects the approval votes to previous blocks. The creator forms a CoSi signature [20] from each set of approval votes and record the signature in the current block. The creator then links backward to the qualified previous block with the largest re-hash value of the previous block hash value.

### *1) The Jump-step Anti-collusion Design*

Note that for the following discussion, we use a single creator case to explain the proposed jump-step algorithm. The validity of the algorithm for the enhanced multiple redundant creators will still hold since a unique block will be selected after each consensus vote.

According to our algorithm, both the future block creator and voters of block $B_{i+2}$ are selected from the $B_i$ block. Therefore, the algorithm is also named the jump-step selection algorithm. Since the maintainers of an even numbered block are always selected from the last even numbered block, and same for odd numbered blocks, we may simply use three bits to record the selection results for each participant. We use one bit to indicated whether one is selected as a maintainer, the second bit to indicate if the one is selected to be a creator or voter, and the third bit to indicate if one is selected to the next even or odd numbered block. For those have been selected will be excluded from being selected again.

The design of separated validation from creation is to avoid collusion and manipulation. The voter selected is to verify the previous block, not current block. Then the creator of the current block uses the verification results to create current block. Additionally, since no one can be selected as a maintainer consecutively, our approach greatly reduces the collusion possibility.

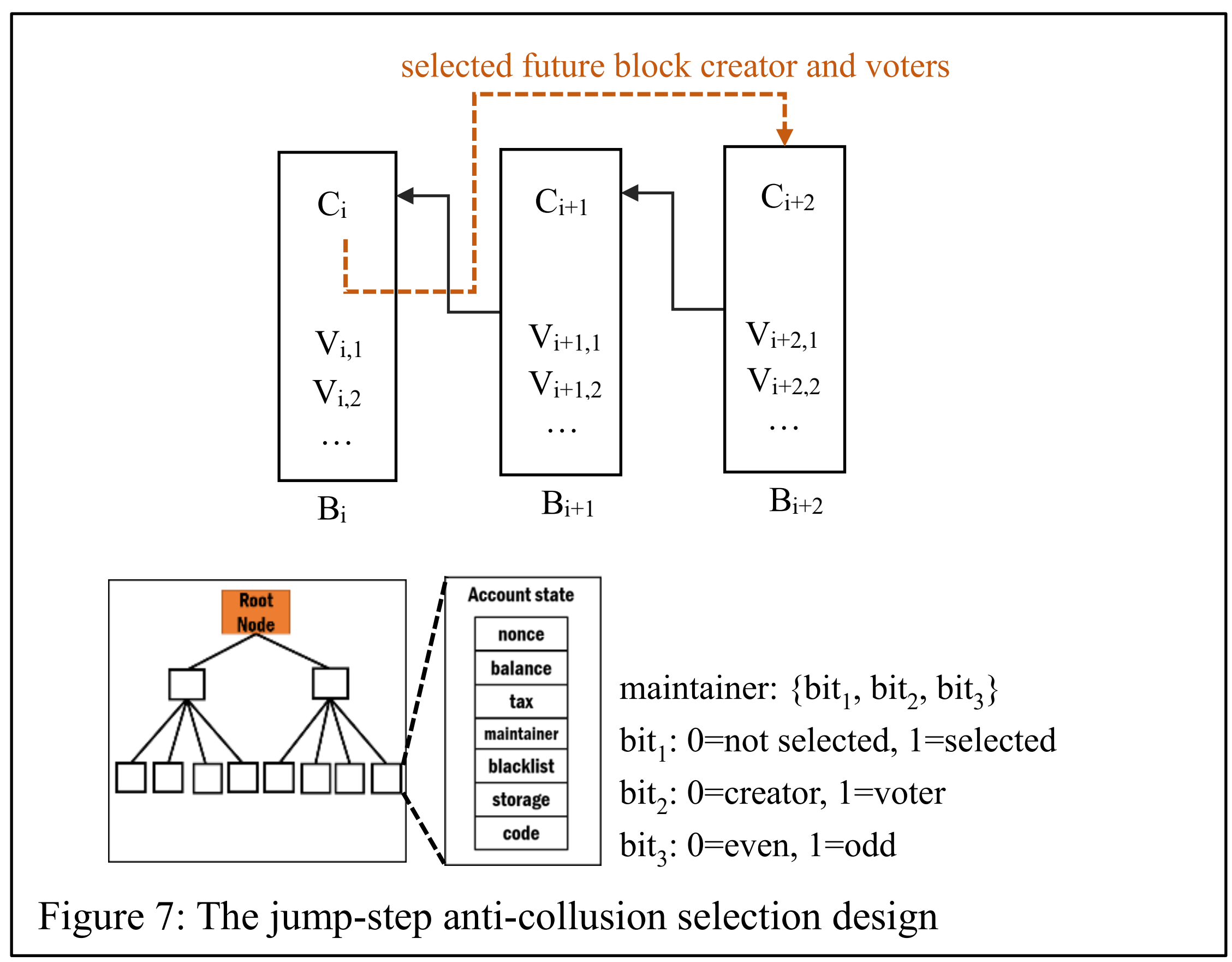

Figure 7: The jump-step anti-collusion selection design

#### *a) Verification process*

For the generation of block $B_{i+2}$, the active creator and the voters are selected in the $B_i$ block. A unique arrangement of our proposed validation approach is to have the current voters approve previous block, i.e. the $B_{i+1}$ block, if the current block is the $B_{i+2}$ block.

Note that the creator $c_{i+2}$ of the $B_{i+2}$ block is selected in the $B_i$ block. Therefore, since voters in the $B_{i+2}$ block is to approve the $B_i$ block, we should choose the voters from at least the $B_i$ block or beyond to avoid the same selected group to perform both block creation and voting.

The selected creator $c_{i+2}$ for the $B_{i+2}$ block validates and accepts pending transactions for current $B_{i+2}$ block. At the same time, the active $B_{i+2}$ voters approve previous block $B_{i+1}$ and issue CoSi signature for the creator to incorporate into current block. When the previous block is approved with CoSi signature of voters, the creator $c_{i+2}$ records in the block header the previous block's hash value to complete the forward link.

One last step to complete the block creation, is to select new creator and voters. The creator $c_{i+2}$ rehashes the concatenated value of the total refundable tax, each active maintainer (creator and voters) address and the maintainer's sequence number to calculate a verifiable and deterministic random number for selection of future maintainer to each corresponding position. Details of the selection process is elaborated next.

***b) Selection process***

The classical consensus methods have the first chain node who solves a given puzzle be the block creator. The beauty is that the operation is autonomous and reliable if offered proper incentive. However, the self-proclaiming winners and endorsers may result in forks and undetermined finality. Some then apply verifiable random function method to select creator and voters.

Now we discuss details of the selection process. First, we assume that the $k$-th maintainer has the account address $a_k$. After the creator finalizes the transactions to be incorporated in the current block, and updates the refundable taxes of nodes affected, on the Merkle Patricia trie, from bottom up, the total refundable tax of each intermedia node is computed from the children nodes. Then on the Merkle root we have the total taxable tax $T$. We then compute

$$h = hash(T||a_k) \bmod T$$

Assume there are $c$ child nodes and the child node $i$ has a total refundable tax $t_i$ collected from its children. We then compute a total left-side-sibling refundable tax $T_i = \sum_{j=1}^{j=i-1} t_j$, and have $T_1 = 0$. The selection then goes to the child $i$ if $0 \leq h - T_i < t_i$. We perform this process recursively down until a leaf node is reached, then the final leaf node reached is the newly selected maintainer. Any node can *efficiently* check the maintainer bits in the Merkle trie to know whether to serve the role of creator or voter. In practice, we exclude the selected maintainers from the selection process to avoid a candidate being consecutively selected.

Note that once each maintainer finishes the assigned task, a fixed reward is granted. At the same time, the maintainer's refundable tax is deducted by the same amount of the award. In this way, we achieve *fairness* by reducing the possibility of being repeatedly selected.

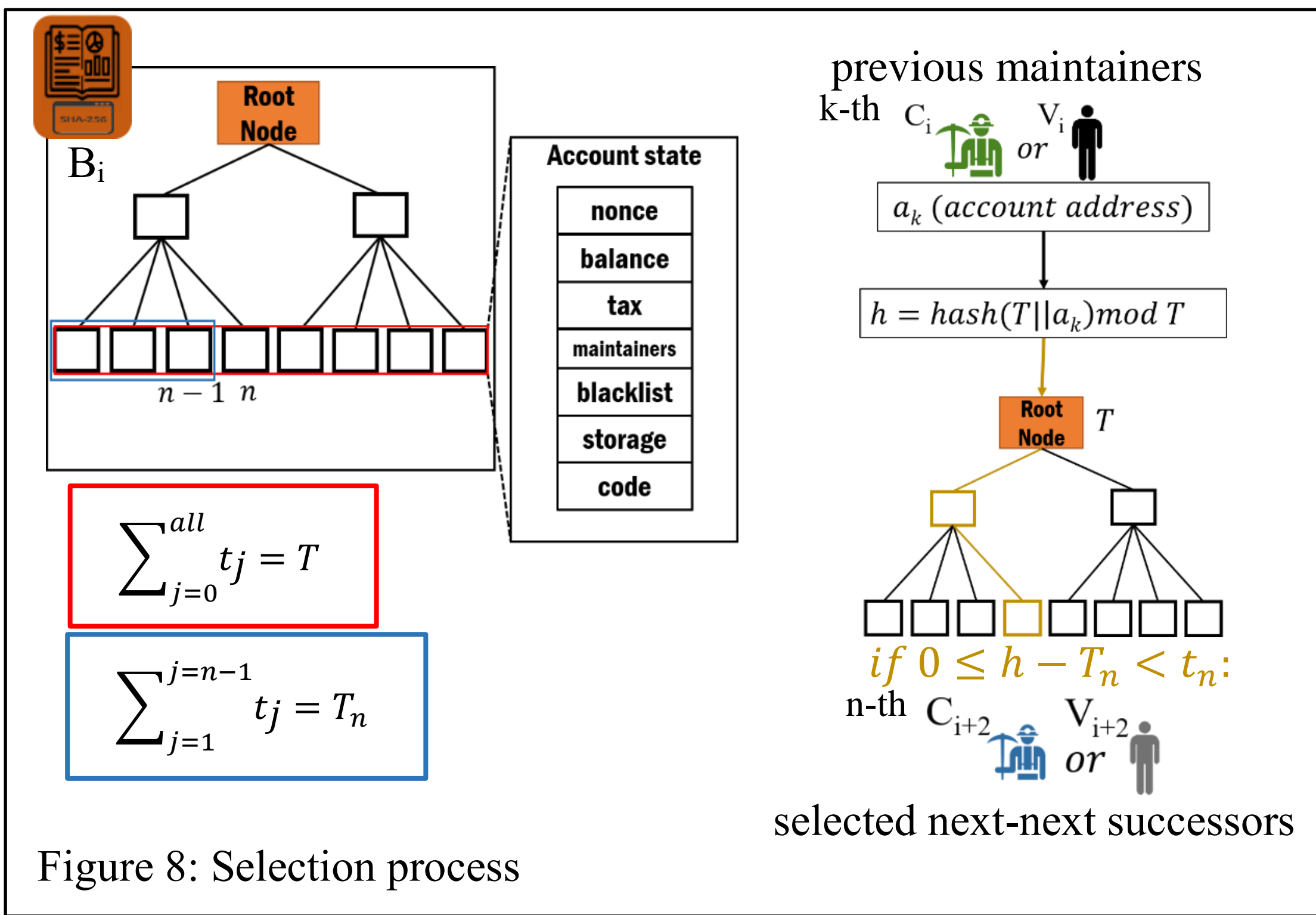

Figure 8: Selection process

## *C.* Refund Mechanism

When each maintainer completes his responsibility, he receives a reward which is then deducted from personal refundable tax. Generally, each individual's refundable tax is accumulated by taxing a fixed percentage of every transaction value. Each block creator is responsible for the tax refund and collection computing.

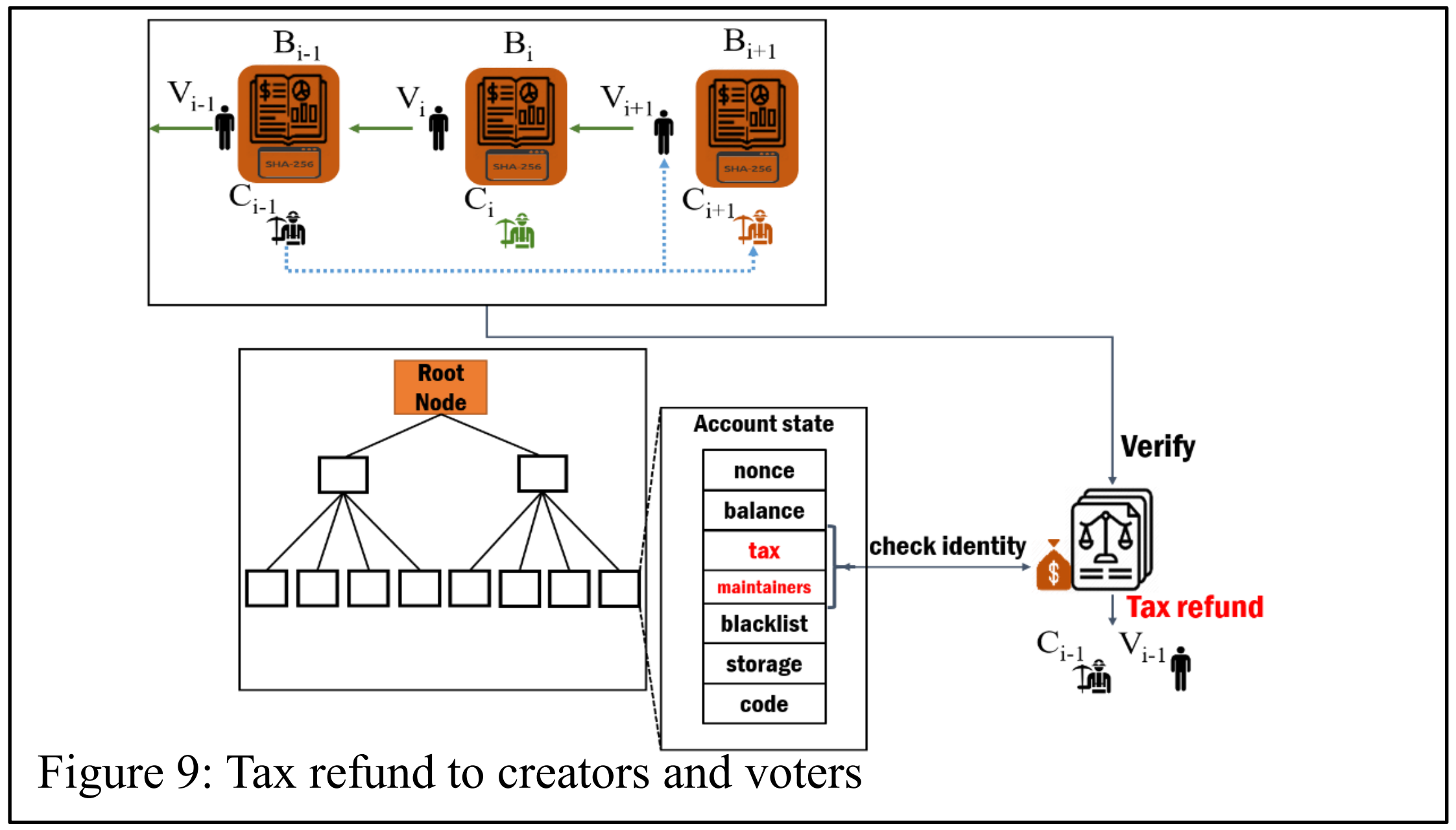

Figure 9: Tax refund to creators and voters

## *D. Blacklist*

Although the PoRT approach is highly secure, we still devise a blacklist scheme to prevent any possible corrupted event. Since the blockchain is public, hackers can attack any time. Nevertheless, since our double-linked PoRT approach creates only one forkless chain with openly verifiable ledger, any corruption can be detected even afterward.

Therefore, we devise a penalty scheme that rewards the whistle blower who reports a fraud validated by creator and approved by voters, while at the same time the involved corruptors are blacklisted and deprived from the privilege of further involvement in the blockchain activities.

At the voting phase, in case any voter identifies a malicious behavior then the voter shall cast a disapproval vote and have the creator blacklists the involved nodes. If necessary, the criminal's account "balance" can be confiscated to compensate for reliability.

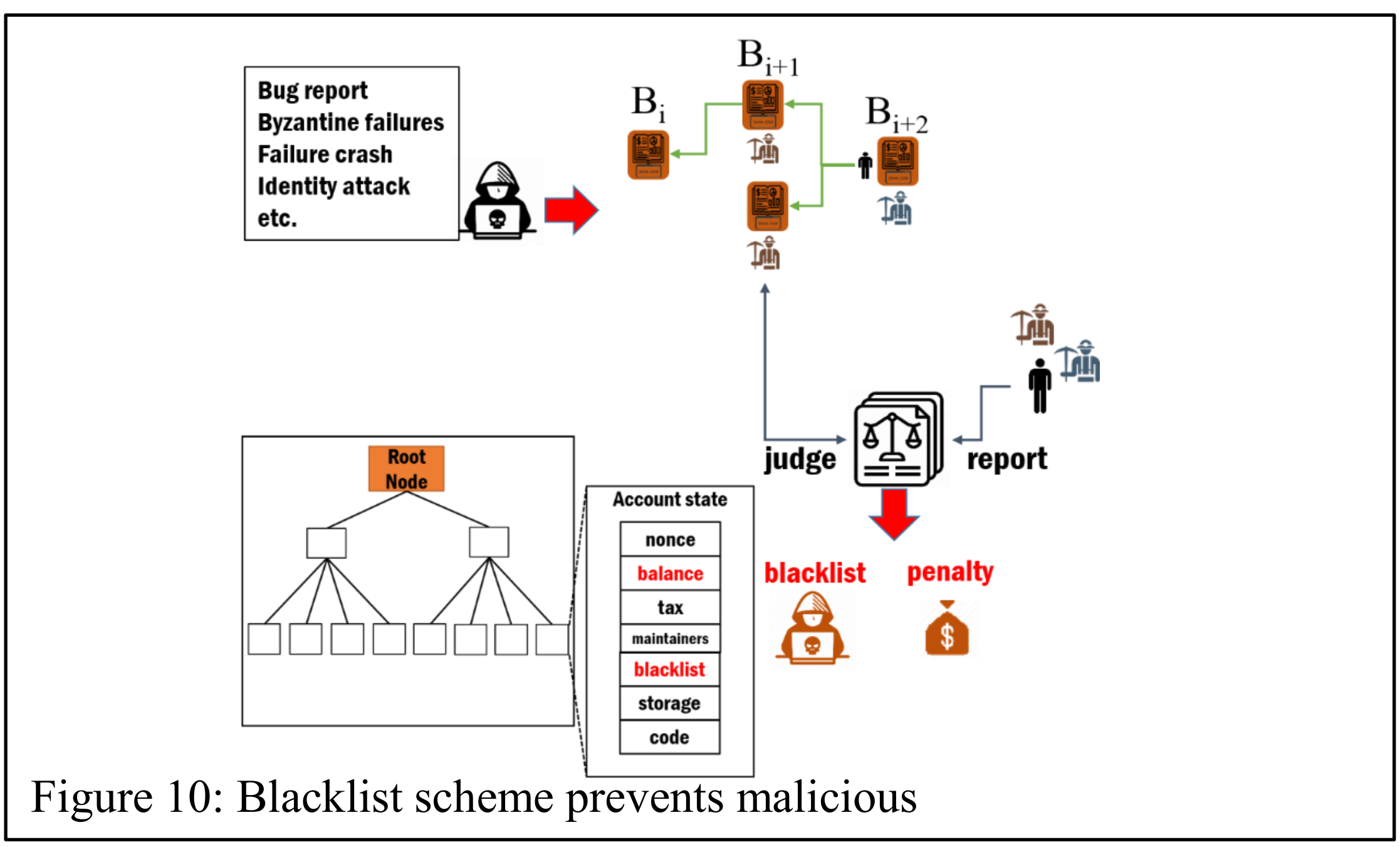

Figure 10: Blacklist scheme prevents malicious

# V. Discussion

This paper proposes a double-linked data structure and an improved consensus mechanism that greatly improves the reliability, efficiency, fairness and stability of blockchain operations. We now discuss a few other issues in designing the blockchain system:

**Incentivized bootstrapping reward.** Since the PoRT approach relies on the refundable tax to select block maintainers, there is an issue at the blockchain startup or genesis phase when no one or few have done transactions and paid taxes. Then no one can be qualified for selection. This issue can be easily resolved, by setting a virtual refundable tax $t_i' = t_i + 1$ as a base for maintainer lottery process. Therefore, in the bootstrapping or genesis phase everyone can still participate in block maintenance and win rewards. If the reward of maintenance is more than the collected refundable tax, then the refundable tax is simply set to zero. Therefore, maintainers can accumulate wealth at the initial phase, while the total wealth will maintain a stable distribution. The PoRT design goal is to have a self-sustained system as the active traders will pay more taxes while they also have higher probability to serve as maintainers to get tax refunded. However, detailed analysis of this self-sustaining stable wealth distribution scheme will need to be further analyzed and validated in the future.

**Security of published maintainer list.** Attackers may attempt to attack newly selected maintainers the current creator publishes. Since our approach proceed very efficiently, the attackers cannot catch up the block creation speed and will fail to have effective attacks.

# VI. Conclusion

We propose a double-linked blockchain based on a proof-of-refundable-tax (PoRT) consensus algorithm and the resulted blockchain achieves high reliability, efficiency, fairness and stability with every computing step publicly verifiable. The most significant advantage of the PoRT approach is the stable wealth distribution and the secure double-linked chain structure which guarantee an unambiguous single chain with no possibility of forks. Therefore, the processing speed is much faster than most existing blockchain designs and good for practical use.